# Sparse-GEV: Sparse Latent Space Model for Multivariate Extreme Value Time Series Modeling

**Yan Liu, Mohammad Taha Bahadori** {YANLIU.CS, MOHAMMAB@USC.EDU}@USC.EDU
Computer Science Department, University of Southern California, Los Angeles, CA 90089 USA

**Hongfei Li** LIHO@US.IBM.COM
IBM T.J. Watson Research Center, Yorktown Heights, NY 10598 USA

## Abstract

In many applications of time series models, such as climate analysis and social media analysis, we are often interested in extreme events, such as heatwave, wind gust, and burst of topics. These time series data usually exhibit a heavy-tailed distribution rather than a Gaussian distribution. This poses great challenges to existing approaches due to the significantly different assumptions on the data distributions and the lack of sufficient past data on extreme events. In this paper, we propose the Sparse-GEV model, a latent state model based on the theory of extreme value modeling to automatically learn sparse temporal dependence and make predictions. Our model is theoretically significant because it is among the first models to learn *sparse* temporal dependencies among multivariate extreme value time series. We demonstrate the superior performance of our algorithm to the state-of-art methods, including Granger causality, copula approach, and transfer entropy, on one synthetic dataset, one climate dataset and two Twitter datasets.

## 1. Introduction

Time series analysis and modeling have been extensively studied in the literature and successfully found applications across domains (Box & Jenkins, 1990; Hamilton, 1994). In many applications, such as climate science, social media analysis and smart grid, we are mostly interested in revealing the temporal dependence and make predictions of extreme events. For example, climate change is mostly characterized by increasing probabilities of extreme weather patterns (IPCC, 2007), such as temperature or precipitation reaching extremely high value. Therefore quantifying the temporal dependence between the extreme events from different locations and make effective predictions are important for disaster prevention; in social media analysis, burst of topics, i.e., buzz, is reflected by extremely high frequency of related words. Uncovering the temporal dependencies between buzzes could reveal valuable insights into information propagation and achieve much better accuracy for buzz prediction.

Identifying temporal dependencies between multiple time-series data is a topic of significant interest (Arnold et al., 2007; Lozano et al., 2009a;b). Many algorithms are proposed to automatically recover the temporal structures, such as autocorrelation, cross-correlations (Box & Jenkins, 1990), randomization test (Edgington & Onghena, 2007), Granger causality (Granger, 1980), transfer entropy (Beirlant et al., 1997; Barnett et al., 2009), and so on. However, uncovering temporal dependency for extreme values is much more challenging than classical observations since the distributions of extreme values are more complex and significantly different from the commonly used Gaussian distribution. In addition, the lack of sufficient past observations on extreme events poses difficulties in modeling and attributing such events.

The statistical approach we can utilize to solve these important problems is the theory of extreme value modeling (Coles, 2001; Beirlant et al., 2004), which provides a natural family of probability distributions for modeling the magnitude of the largest (or smallest) of a large number of events, and a canonical stochastic process model (Coles, 2001) for the occurrence of events above a very high (or below a very low) threshold. In the past decade, extreme value modeling has attracted a lot of research efforts in statistics, finance, and environmental science, particularly on modeling temporal and spatio-temporal extreme value (Coles & Tawn, 1996; Ferro & Segers, 2003; Huerta & Sanso, 2007). However, all of the work above model temporal or spatial dependence with predefined covariance structures (e.g. without independence considerations). Furthermore, most general discussions of dependencies in multivariate extreme value modeling has been focused on pairwise relationships. This is obviously unrealistic and demands a significant contribution on automatically learning the temporal structures from the data for better analysis and modeling.

In this paper, we propose a sparse latent space model, namely sparse-GEV model, to solve the problem. The basic idea of our approach is to model the multivariate extreme value time series as a latent space model. The latent variables, corresponding to the location parameters (which determine the mode) of extreme value distributions for time series at certain time, are modeled by the location parameters of all time series in history, through a dynamic linear model (DLM). By imposing an $L_1$-penalty with respect to the regression coefficients in DLM, we could establish meaningful temporal dependencies between a small subset of time series and the concerned time series of extreme values. To estimate parameters of the model, we develop a iterative searching algorithm based on the generalized EM-algorithms and sampling with particle filtering. Our model is significant because it is among the first models to reveal the temporal dependencies between multiple extreme value time series. In addition, our experiment results demonstrate the superior performance of our model to other state-of-art methods on both learning temporal dependence and predicting future value.

The rest of the paper is organized as follows: we first describe the details of our proposed model in Section 2, then we review the existing work and discuss their connections to our model in Section 3, we show the experiment results in Section 4, and finally we summarize the paper with hints on future work.

## 2. Methodology

**Preliminaries** Before diving into the details of our model, we first briefly review the extreme value theory (Coles, 2001). Let $X_1, \cdots, X_m$ be a sequence of independent and identically distributed random variables, and let $M_m = \max\{X_1, \cdots, X_m\}$. If there exist sequences of constants $a_m > 0$ and $b_m$ such that

$$\Pr\left(\frac{M_m - b_m}{a_m} \leq z\right) \to G(z) \quad \text{as } m \to \infty, \quad (1)$$

for some non-degenerate distribution function $G$, then $G$ should belong to the generalized extreme value (GEV) families, namely

$$G(z) = \exp\left\{-\left[1 + \xi\left(\frac{z - \mu}{\sigma}\right)\right]_+^{-1/\xi}\right\}, \quad (2)$$

defined on $\{z : 1 + \xi(z - \mu)/\sigma > 0\}$, where $\mu$ ($-\infty < \mu < \infty$) is the location parameter, $\sigma$ ($\sigma > 0$) is the scale, and $\xi$ ($-\infty < \xi < \infty$) is the shape parameter $\xi$, which governs the tail behavior of the distribution.

One popular GEV distribution is the Gumbel distribution when $\xi \to 0$, whose pdf is defined as

$$p(z|\mu, \sigma) = \frac{1}{\sigma} \exp\left\{-\frac{z - \mu}{\sigma} - \exp\left(-\frac{z - \mu}{\sigma}\right)\right\}. \quad (3)$$

It has been shown that the maximum value in a sample of a random variable following an *exponential family distribution* (such as Gaussian, Lognormal and Gamma distributions) converge to the Gumbel distribution. One special property of the Gumbel distribution is that the mode is determined solely by the location parameter $\mu$.

### 2.1. Model Description

Given multivariate time series data, our goal is to build an effective model that can recover temporal dependence between extreme value time series (block maxima or peaks over threshold) and make accurate predictions for future extreme events. To achieve a robust and interpretable model, a natural choice is to capture the temporal dependence via linear models; however, this is not directly achievable on extreme value variables since their temporal dependence is obviously nonlinear. To solve the problem, we propose latent models in which the location parameters of GEV distributions are latent variables and the temporal dependence between extreme value variables is captured via the latent variables through dynamic linear model. We choose the location parameters because they capture the mode of extreme value variables and can be modeled reasonably well by linear dependence.

Formally, let $\boldsymbol{x}^1, \ldots, \boldsymbol{x}^P$ denote P number of extreme value time series and each time series $\boldsymbol{x}^i$ have $T$ observations, i.e., $\boldsymbol{x}^i = \{x_1^i, \ldots, x_T^i\}$[1], we define the joint

---

[1] In extreme value theory, two main sets of methods, the Block Maxima method and the Peaks over Thresholds method have been developed to model extreme values (Coles, 2001). In the rest of the paper, we use Block Maxima method as an example to describe our model. Notice that our methodology is applicable to the Peaks over Thresholds approach by defining a point process model and in the experiments we have used both approaches.

probability of observations $\{x_t^i\}$ and their associated location parameter $\{\mu_t^i\}$ as:

$$p(\{x_t^i\}, \{\mu_t^i\}|\boldsymbol{\beta}, \boldsymbol{\sigma}, \boldsymbol{c}) = \prod_{i=1}^{P}\prod_{t=L+1}^{T} p(x_t^i|\mu_t^i, \sigma^i) p(\mu_t^i|\{\mu_{t-l}^j\}, \boldsymbol{\beta}, \boldsymbol{c}), \quad (4)$$

where $p(x_t^i|\mu_t^i, \sigma^i)$ can be modeled by a GEV distribution such as the Gumbel distribution in eq(3) with $\sigma^i$ as the scale parameter specific to time series $x^i$, $\{\mu_{t-l}^j\}$ is the history of all time series at time $t$ with a maximal lag of $L$, and $p(\mu_t^i|\{\mu_{t-l}^j\}, \boldsymbol{\beta}, \boldsymbol{c})$ can be modeled by a dynamic linear model as follows,

$$\mu_t^i = c^i + \sum_{l=1}^{L}\sum_{j=1}^{P} \beta_{j,l}^i \mu_{t-l}^j + \epsilon. \quad (5)$$

where $c^i$ is the offset specific to time series $i$, $\boldsymbol{\beta}^i$ are the coefficients, and $\epsilon$ is a Gaussian noise with variance $\tau^2$. As we can see, the temporal dependence between $\boldsymbol{x}^i$ and $\boldsymbol{x}^j$ is now captured via the coefficients $\boldsymbol{\beta}$. By adding a shrinkage Laplace prior over $\boldsymbol{\beta}$ when maximizing the likelihood function, i.e.,

$$\{\hat{\boldsymbol{\beta}}, \hat{\boldsymbol{\sigma}}, \hat{\boldsymbol{c}}\} = \arg\max \ell(\boldsymbol{x}^1, \ldots, \boldsymbol{x}^P; \boldsymbol{\beta}, \boldsymbol{\sigma}, \boldsymbol{c}) + \sum_{i=1}^{P} \lambda \|\boldsymbol{\beta}^i\|_1, \quad (6)$$

where $\lambda$ is the regularization parameter, we can obtain the sparse solution of $\beta$. Finally, we determine that $\mathbf{x}^i$ temporally depends on $\mathbf{x}^j$ if the corresponding value of $\boldsymbol{\beta}_j^i$ is non-zero. In this way, our model not only can provide better understanding of potential causes of the extreme events, but also helps to achieve more accurate prediction of the extreme events in the future. This model is later referred to as the *Sparse-GEV* model.

### 2.2. Inference and Learning

Given the existence of hidden variables in the sparse-GEV model, directly maximizing the likelihood as in eq(6) is not feasible. Therefore we applied the generalized EM-algorithm to solve the problem.

Next, we use Gumbel distribution as an example to demonstrate how we can make efficient inference and learning in the proposed model. In the EM algorithm, we optimize the following function via two steps:

$$Q(\boldsymbol{\beta}, \boldsymbol{\sigma}, \boldsymbol{c}; \boldsymbol{\beta}^{old}, \boldsymbol{\sigma}^{old}, \boldsymbol{c}^{old}) = -\sum_{i=1}^{P}\sum_{t=L+1}^{T} \ln(\sigma^i)$$
$$- \mathbb{E}_{\{\boldsymbol{\mu}|\boldsymbol{X},\boldsymbol{\beta}^{old},\boldsymbol{\sigma}^{old},\boldsymbol{c}^{old}\}} \left[ \frac{x_t^i - \mu_t^i}{\sigma^i} + \exp\left(-\frac{x_t^i - \mu_t^i}{\sigma^i}\right) \right.$$
$$\left. + \frac{1}{2}\left(\frac{\mu_t^i - c^i - \sum_{l=1}^{L}\sum_{j=1}^{P}\beta_{j,l}^i \mu_{t-l}^j}{\tau}\right)^2 \right]. \quad (7)$$

**E Step** Directly calculating the expectations in eq(7) is infeasible given the form of the posterior probability, therefore we apply sampling algorithms for approximation. In order to generate samples from $p\{\boldsymbol{\mu}|\boldsymbol{X}, \boldsymbol{\beta}^{old}, \boldsymbol{\sigma}^{old}, \boldsymbol{c}^{old}\}$, we use the particle filtering algorithm (Doucet & Johansen, 2009). The major challenge is that in each iteration of particle filtering, we need to draw samples from $p(\mu_t^i|x_t^i, \{\mu_{t-l}^j\})$, which cannot be calculated analytically. Instead, we use the following proposal function:

$$\mathcal{N}\left(\tilde{\mu}_t^i + \gamma_i \tau - \sigma^i \mathcal{W}_0\left(\gamma_i^2 \exp\left(\frac{\tilde{\mu}_t^i - x_t^i}{\sigma^i} + \gamma_i^2\right)\right), \frac{\tau^2}{\gamma_i^2 + 1}\right),$$

where $\mathcal{W}_0$ is the Lambert W function, $\gamma_i = \tau/\sigma^i$, and $\tilde{\mu}_t^i$ is calculated using the history, i.e., $\tilde{\mu}_t^i = c^{i,old} + \sum_{l=1}^{L}\sum_{j=1}^{P} \beta_{j,l}^{i,old} \tilde{\mu}_{t-l}^j$. The rationale behind this choice is to approximate the posterior distribution $p\{\boldsymbol{\mu}|\boldsymbol{X}, \boldsymbol{\beta}^{old}, \boldsymbol{\sigma}^{old}, \boldsymbol{c}^{old}\}$ with a Gaussian distribution with the same mode and similar variance.

Notice that particle filtering may encounter the challenge of "miniscule weights" if the sequence length is long. Therefore the resampling step is usually applied at each time stamp to resolve the issue (Doucet & Johansen, 2009). For very long time series, particle filtering does face some other challenges, but can be fixed using particle smoothing (Doucet & Johansen, 2009).

**M Step** The optimization problem for updating $\boldsymbol{\beta}^i$ and $c^i$ is as follows:

$$\min_{\boldsymbol{\beta}^i, c^i} \mathbb{E}_{\boldsymbol{\mu}|\mathbf{x}} \sum_{t=L+1}^{T} \left(\mu_t^i - c^i - \sum_{l=1}^{L}\sum_{j=1}^{P} \beta_{j,l}^i \mu_{t-l}^j\right)^2 + \lambda \|\boldsymbol{\beta}^i\|_1,$$

where the expectation is computed from the samples. As we can see, the optimization function has the Lasso format and can be solved efficiently by algorithms such as coordinate descent (Wu & Lange, 2008). The parameter estimation for the Gumbel distribution itself is not a trivial problem. In general, the MLE is the widely accepted approach to estimate the shape and scale parameters, and Newton-Raphson or quasi-Newton methods can be applied to solve the resulting optimization problem (Evans et al., 2000). Therefore we estimate $\boldsymbol{\sigma}$ by the Newton-Raphson algorithm.

### 2.3. Prediction

In order to make predictions on the future value of extreme events, for example $x^i_{T+1}$, given the extreme value time series up to time $T$, we can first estimate the mean $\bar{\mu}^i_{T+1}$ using the samples drawn from the posterior distribution with the learned parameters. Based on the model defined in eq(4), we can predict $x^i_{T+1}$ as

$$\hat{x}^i_{T+1} = \bar{\mu}^i_{T+1} + \gamma_E \sigma^i,$$

where $\bar{\mu}^i_{T+1} = c^i + \sum_{l=1}^{L}\sum_{j=1}^{P} \beta^i_{j,l} \bar{\mu}^j_{T-l+1}$, and $\gamma_E (\approx 0.5771)$ is the Euler constant.

### 2.4. Scalability

The computational complexity of Sparse-GEV depends on two factors: the number of EM iterations required for convergence and the scalability of E-Step and M-Step. In Section 5, we empirically show that EM usually converges within a small number of iterations. In the M-Step, while there are efficient solvers for both equations, the problems for different time series are independent and can be implemented in parallel. The particle filtering in E-Step is notoriously efficient for sampling from time series mainly due to three reasons: (i) it requires only one iteration to generate the samples, (ii) the generated samples are independent; no burn-in period or decoupling is required and (iii) at each time stamp the sampling procedures in different locations are independent from each other and can be implemented in parallel. Therefore our algorithm is scalable and could be easily applicable to practical applications.

## 3. Related Work and Discussions

Very recently, a few advanced approaches have been explored to uncover temporal dependence from time series data, including Lasso-Granger (Arnold et al., 2007), transfer entropy (Schreiber, 2000), and the copula approach (Liu et al., 2009). In this section, we discuss how these algorithms can be applicable to extreme value time series analysis and their connections to Sparse-GEV.

### 3.1. Related Work

**Granger causality** In (Arnold et al., 2007), the Lasso-Ganger algorithm, an effective and efficient approach to learn sparse temporal graphs, is developed by combining Granger causality with sparse neighborhood selection using $L_1$-penalized regression. More specifically, given $p$ number of time series, $\mathbf{x}^1, \ldots, \mathbf{x}^p$, where $\mathbf{x}^i = \{x^i_t : t = 0, \ldots, T\}$, let $X^{Lagged}_{t,L}$ represent the concatenated vector of all the lagged variables (with a maximal lag of L) of up to time $t$, i.e., $\{x^{t-l}_j : j = 1, \ldots, p, l = 1, \ldots, L\}$. Then the temporal graphs can be learned by the following regularized regression:

$$\hat{\boldsymbol{\beta}}_i(\lambda) = \arg\min_{\beta_i}(\sum_{t=1}^{T} \|x^t_i - X^{Lagged}_{t,L}\boldsymbol{\beta}_i\|^2 + \lambda\|\boldsymbol{\beta}_i\|_1), \quad (8)$$

where there is an edge from $\mathbf{x}^j$ to $\mathbf{x}^i$ if and only if at least one of the corresponding coefficients in $\hat{\boldsymbol{\beta}}_i$ is non-zero. The Lasso-Granger algorithm can be directly applied to extreme value observations to infer the dependency graph, but obviously this violates the common assumptions of linear dependency in Granger causality.

**Transfer Entropy Solution** Transfer entropy is usually employed when the data do not follow the autoregressive model and a nonlinear generalization of the Granger causality framework is desirable. In the Transfer entropy framework (Schreiber, 2000), time series $\boldsymbol{x}^i$ is thought to be a cause of another time series $\boldsymbol{x}^j$ if the values of $\boldsymbol{x}^i$ in the past significantly decrease the uncertainty in the future values of $\boldsymbol{x}^j$ given its past. The amount of decrease in the uncertainty can be quantified as

$$T_{\boldsymbol{x}^i \to \boldsymbol{x}^j} = H(x^j_t | x^i_{t-L:t-1}) - H(x^j_t | x^j_{t-L:t-1}, x^i_{t-L:t-1}),$$

where $H(x)$ is the Shannon entropy of the random variable $x$. Since the transfer entropy is a pairwise quantity, we can use its output as input to a graph learning algorithm, for example, IAMB (Tsamardinos et al., 2003), to uncover the temporal dependency among multiple time series.

The transfer entropy approach can be used to uncover causality relationship among extreme value time series since it does not rely on any particular assumptions on the distribution of the time series.

**Copula Approach** The copula approach has been proposed for dependency analysis of time series with non-Gaussian marginal distributions (Embrechts et al., 2002). It has been used for forecast in time series (Leong & Valdez, 2005) and learning sparse dependency structures (Liu et al., 2009). In a copula framework, e.g., Gaussian copula, the marginal distribution of the time series $\mathbf{X}^i$ are estimated as $\tilde{F}_i$. Next the observations are transformed to the Gaussian copula domain as $U^i_t = \Phi^{-1}(\tilde{F}_i(X^i_t))$, where $\Phi$ is the cdf of the unit Gaussian distribution. Finally the temporal causal graph can be uncovered by analysis of dependency among $U^i_t$ using algorithms such as glasso algorithm (Friedman et al., 2008). We report an edge from node $i$ to node $j$ if the precision matrix has at least one non-zero element from lagged $U_{t-\ell}^j$ to $U_t^i$, for $\ell \geq 1$. The method in (Leong & Valdez, 2005) can be used for predicting the future values of the time series.

In order to uncover temporal dependencies among extreme value time series, we can either estimate the marginals with a non-parametric density estimator or use the GEV distribution to estimate the marginal distribution. For extreme value time series, the latter is preferred since the non-parametric approximation of the marginal distributions could lead to over-fitting when the number of observations is scarce.

### 3.2. Connections to existing algorithms

The connections between our algorithm with existing algorithms can be established by considering Sparse-GEV, transfer entropy and the copula approach as extensions of the Granger causality framework. The copula approach leverages the marginal distribution of the time series to map the observations to another space and assumes linear dependence in the new space. Sparse-GEV discovers the Granger causality relationship among the latent variables from which the observations have been generated. The transfer entropy approach generalizes the Granger causality framework by finding the Granger causality type relationships from the uncertainty of the time series. In fact, when the data are distributed according to Gaussian linear model, transfer entropy is equivalent to Granger causality (Barnett et al., 2009).

For high-dimensional time series, the number of observations is much less than the parameters of the model. The Lasso-Granger algorithm benefits from the variable selection properties of Lasso. (Meinshausen & Bühlmann, 2006) show that the Lasso variable selection loss, and subsequently the Lasso-Granger's loss (Arnold et al., 2007), vanishes with an exponential rate. For the copula approach, (Liu et al., 2009) show that when copula-based model is the true model, the copula-based structure learning algorithm with non-parametric estimation of marginals converges to the true graph with a rate of $\mathcal{O}\left(\sqrt{\frac{\log(n)}{n^{1-\xi}}}\right)$ for some $\xi \in (0, 1)$, which is far slower than the exponential convergence of Lasso-Granger. The performance of transfer entropy heavily relies on the accuracy of entropy estimations, which require a large number of observations, especially for high dimensional distributions, to achieve robust estimation (Beirlant et al., 1997). For example, the Nearest Neighbor Estimator converges with root-n rate, which is again far slower than the convergence rate of Lasso-Granger. However, Sparse-GEV inherits the variable selection advantages of Lasso-Granger while allows a more flexible marginal distribution for the observations. It is fully parametric, and together with proper $\ell_1$ penalization can avoid over-fitting while capturing non-linear dependencies.

## 4. Experiment Results

In order to evaluate the effectiveness of our algorithm, we conduct experiments on four datasets, including one synthetic dataset, one weather dataset and two Twitter datasets. The experiment results are evaluated on both how well we uncover the temporal dependence graphs and how accurately we can predict the future value of extreme events using the learned temporal dependence.

### 4.1. Datasets

**Synthetic Dataset** We generate eight synthetic datasets, each composed of nine time-series with different types of temporal dependence, one of which is shown in Figure 1(a). Time series of length $T = 40$ are generated in two steps: (i) A set of observations of the location variables $\tilde{\boldsymbol{\mu}}$ is generated according to eq (4), with the offset $\{c^i\}$ generated from $N(0.2, 0.05)$, the coefficients $\boldsymbol{\beta}$ set to have stationary time series, $\tau^2$ set to 0.1 and the time lag $L$ set to 2; (ii) The observations $\tilde{\boldsymbol{x}}$ are generated from a Gumbel distribution with the corresponding location parameters $\tilde{\boldsymbol{\mu}}$ and scale parameter $\sigma^i = 0.05$ for all time series.

**Climate Dataset** The study of extreme value of wind speed and gust speed is of great interest to the climate scientists and wind power engineers. A collection of wind observations is provided by AWS Convergence Technologies, Inc. of Germantown, MD. It consists of the observations of surface wind speed (mph) and gust speed (mph) every five minutes. We choose 153 weather stations located on a grid laying in the $35N - 50N$ and $70W - 90W$ block. Following the traditions in this domain, we generated extreme value time series observations, i.e, daily maximum values, at different weather stations. The objective is examine how the wind speed (or gust speed) at different locations affects each other and how well we can make predictions on future wind speed.

**Twitter Dataset** In social media analysis, "buzz" refers to those topics or memes that many people are talking about at the same time with rapid growth and impact. Buzz modeling and predictions are the fundamental problems in computational social science, but they are extremely challenging since the distributions of these time series observations have heavy tails and most existing models fail miserably. Given the definition of buzz, i.e., extremely high frequency of certain words within a time interval, it is natural to model them via extreme value theory. We collected two Twitter datasets to evaluate the effectiveness of our model: one is the most popular 20 meme phrases during a 28-day interval from Nov-Dec 2009, and the other is popular hashtags around "occupy wall street" during a 21-day interval in Oct-Nov 2011. Some example phrases in the first dataset are "Haiti earthquake", "Grammy Awards", "iPad release", and "Scott Brown's Senate Election"; some example hashtags in the second dataset are #OWS, #OccupyLA, #OccupySF, #OccupyDC and #OccupyBoston. For those phrases and hashtags, we count the number of mentions in tweets within a interval of one hour. We are interested in uncovering how different buzzes affect each other and how well we can make predictions on future buzz.

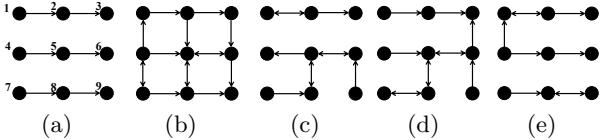

(a)    (b)    (c)    (d)    (e)

Figure 1. Illustration of (a) Ground truth and the inferred temporal dependence graphs by (b) Granger causality, (c) Transfer entropy, (d) Copula method, and (e) Sparse-GEV.

Table 1. Comparison of different models on recovering the temporal dependence graph on eight synthetic datasets.

| Algorithms | Avg AUC Score |
| --- | --- |
| Sparse-GEV | **0.9257** |
| Granger | 0.9046 |
| Transfer Entropy | 0.8701 |
| Copula | 0.8836 |

### 4.2. Performance Comparison

We compare the performance of our Sparse-GEV model with three baselines, including Granger causality, transfer entropy, and the copula method (with Gaussian copula), on two tasks: uncovering the underlying dependency among time series, and predicting the future values of time series. The first one requires knowledge about the true dependency structure, which is only available in the synthetic dataset. For evaluation, we use the Area Under the Curve (AUC) score, i.e., the probability that the algorithm will assign a higher value to a randomly chosen positive (existing) edge than a randomly chosen negative (non-existing) edge in the graph. In the prediction task, we conduct experiments via the sliding window approach: given time series observations of length $T$ and a window size $S$, we train a model on observations of $x_s, \ldots, x_{T-S+s-1}$ and test it on the $(T-S+s)^{th}$ sample, for $s = 1, \ldots, S$. We set $S$ to 10 for all the datasets and use the root mean squared error (RMSE) measure (averaged over $S$ experiments and all nodes) as the evaluation metric.

In the experiment, the regularization parameter $\lambda$ is set via cross-validation. All the observations are normalized into interval $[0, 1]$ prior the experiments.

**Temporal Dependence Discovery** Table 1 lists the average accuracy of uncovering the underlying dependence structures by different algorithms on the synthetic data (consisting of 8 different datasets). As we can see, our Sparse-GEV model significantly outperforms the baseline methods. Figure 1 shows an example of the graphs learned by different algorithms: our model can recover the ground-truth graph more accurately than other methods.

Fig. 2 shows the inferred temporal dependencies from the extreme value time series of wind speed and wind gust speed by Sparse-GEV. Given the limited space, we limit our discussion on the new york region. The main observation is that the weather in the inland regions are heavily influenced by the coastline region. The wind gust graph (Fig. 2(b)) indicates two clusters. One is at the top of the graph, starting from Middletown to Danbury across Fishkill. The other one is located at the bottom of the graph, which passes through several cities, such as Stamford, Fairfield and Brookhaven, around Long Island Sound, then goes to inland cities in New Jersey through New York City. The top cluster gives an example of wind gust path in inland while the bottom one shows the coastal impact of Long Island Sound and the impact extends to inland New Jersey. Comparably, in addition to the Middle-

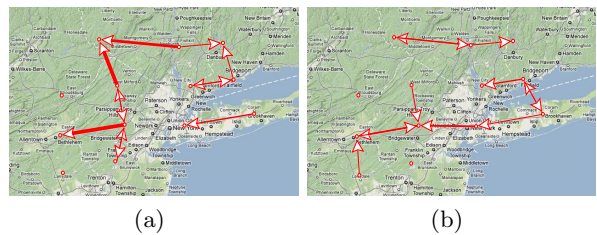

(a)    (b)

Figure 2. The temporal dependence graph learned by Sparse-GEV on the extreme value time series of (a) Wind in NY and (b) Gust in NY. Thicker edges imply stronger dependency.

Sparse-GEV: Sparse Latent Space Model for Multivariate Extreme Value Time Series Modeling

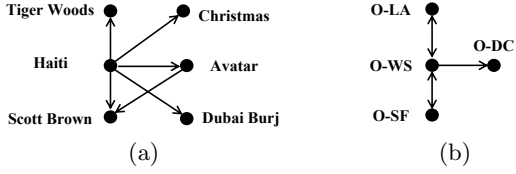

(a)      (b)

Figure 3. The temporal dependency graph learned by Sparse-GEV from the Twitter dataset on (a) Meme phrases in 2009 and (b) "Occupy Wall Street" hastags in 2011.

town to Danbury inland cluster in wind gust graph, the wind graph (Fig. 2(a)) shows another inland cluster centered at Bridgewater, which has strong temporal dependencies with neighboring cities (confirmed by the climatologists).

Fig. 3 shows the inferred temporal dependencies from the extreme value time series of Twitter data on a subset of buzz. From Fig. 3(a), we can see that the temporal dependence between different buzz are sparse (since they are quite different topics); however, the buzz on "Haiti Earthquake" generates a huge impact on the whole Twitter universe and significantly changes the future mentions of other popular meme phrases. An interesting observation in Fig. 3(b) is that the hashtag on the general theme #OWS has direct temporal dependence with the city-specific hashtags, such as #O-LA and #O-DC, while city-specific hashtags do not affect each other.

**Prediction Performance** As discussed in Section 2, Sparse-GEV can also be used for predicting future extreme events. For other baseline methods, we use the approaches discussed in Section 3.1 for predictions. Table 2 shows the prediction accuracy of different algorithms on all datasets. As we can see, Sparse-GEV outperforms all the other algorithms across all datasets. This can be attributed to two properties of Sparse-GEV: its flexibility in modeling complex distributions and its effectiveness in utilizing the samples. The assumptions of Lasso-Granger and copula methods about the distribution of the data can be responsible for their lower performance. Transfer entropy re-quires many observations to perform well, which could be a potential issue in the real applications.

Table 2. Comparison of RMSE by different in the prediction tasks. TE: transfer entropy; T-: Twitter dataset.

|            | Synth. | Wind  | Gust  | T-Meme | T-OWS |
|------------|--------|-------|-------|--------|-------|
| Sparse-GEV | **.2644** | **.0660** | **.0927** | **.0503** | **.1190** |
| Granger    | .2923  | .0695 | .0943 | .0619  | .1410 |
| TE         | .3135  | .0692 | .0983 | .0972  | .1302 |
| Copula     | .2987  | .0678 | .0934 | .1009  | .1240 |

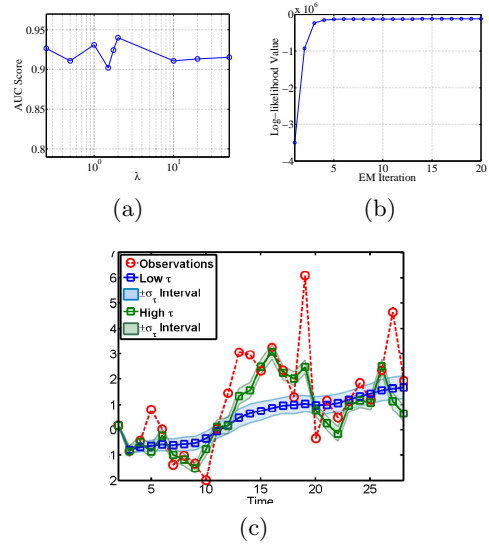

(a)      (b)

(c)

Figure 4. (a) Parameter Sensitivity Assessment: Average AUC achieved by Sparse-GEV on the synthetic datasets when the value of $\lambda$ varies. (b) The value of log-likelihood function at each iteration of the EM algorithm. (c) The effect of $\tau$ on the value of hidden variables in the Sparse-GEV algorithm.

### 4.3. Parameter Sensitivity Assessment

Like other latent state models, Sparse-GEV model has many parameters, which could affect its performance significantly. In our last experiment, we assess the parameter sensitivity. Fig. 4(a) shows that in a large range of values for the regularization parameter $\lambda$, the graph learning accuracy remains unchanged and little effort in selection of the regularization parameter leads to the optimal performance. Fig. 4(b) suggests that in less than 10 EM iterations, our algorithm converges to the optimal point. Fig. 4(c) illustrates the effect of $\tau$ on the performance of Sparse-GEV. Small values of $\tau$ result in smoother estimation of $\mathbb{E}[\boldsymbol{\mu}|\mathbf{X}]$, while higher values lead to sensitive estimation (as a result $\mathbb{E}[\boldsymbol{\mu}|\mathbf{X}]$ closely follows the observation time series). This observation suggests that we should monitor the sample mean of the latent variables and choose a value of $\tau$ that allows smooth latent variables to capture the trend of observations.

## 5. Conclusion

In this paper, we propose sparse-GEV, a sparse latent space model, to uncover the sparse temporal de-pendency from multivariate extreme value time series. To estimate the parameters of the model, we develop an iterative searching algorithm based on the generalized EM-algorithm and sampling with particle filtering. Through extensive experiments, we demonstrate that Sparse-GEV outperforms the state-of-the-art algorithms such as copula and transfer entropy. For future work, we are interested in the theoretical analysis on the consistency of the Sparse-GEV model.

# Acknowledgement

We thank the anonymous reviewers for their valuable comments. This research was supported by the NSF research grants IIS-1134990.